\documentclass[sigconf]{acmart}

\usepackage{booktabs} % For formal tables
\usepackage{graphicx}
\usepackage{textcomp}
\usepackage{xcolor}
\usepackage{comment}
\usepackage{braket}
\usepackage{hyperref}
\usepackage{float}
\usepackage{multirow}
\usepackage{makecell}
\usepackage{bm}
\usepackage{subfig}
\usepackage{setspace}
\usepackage{enumitem}
\usepackage{algorithm}
\usepackage{algpseudocode}
\usepackage{pifont}

\AtBeginDocument{%
  }

\setcopyright{acmlicensed}
\copyrightyear{2018}
\acmYear{2018}
\acmDOI{XXXXXXX.XXXXXXX}
\acmConference[Conference acronym 'XX]{Make sure to enter the correct
  conference title from your rights confirmation email}{June 03--05,
  2018}{Woodstock, NY}
\acmISBN{978-1-4503-XXXX-X/2018/06}

\definecolor{commentColor}{HTML}{6A9F3D}
\newcommand{\com}[2][commentColor]{{\color{#1} #2}}

\begin{document}

\title{Large-Scale Quantum Circuit Simulation on HPC Cluster via\\ Cache Blocking, Boosting, and Gate Fusion Optimization}

\author{Chuan-Chi Wang}
\affiliation{%
  \institution{National Taiwan University}
  \city{Taipei}
  \country{Taiwan}
}
\email{d10922012@ntu.edu.tw}

\author{Yan-Jie Wang}
\affiliation{%
   \institution{National Taiwan University}
   \city{Taipei}
   \country{Taiwan}
}
\email{yanjiewtw@gmail.com}

\author{Chia-Heng Tu}
    \affiliation{%
    \institution{National Cheng Kung University}
    \city{Tainan}
    \country{Taiwan}
}
\email{chiaheng@ncku.edu.tw}

\author{Shih-Hao Hung}
\affiliation{%
    \institution{National Taiwan University}
    \city{Taipei}
    \country{Taiwan}
}
\email{hungsh@csie.ntu.edu.tw}

\begin{abstract}
Quantum circuit simulation is crucial for the development of quantum algorithms, particularly given the high cost and noise limitations of physical quantum hardware. While full-state quantum circuit simulation is commonly employed for prototyping and debugging, it poses challenges because of the exponential increase in simulation time for large quantum systems.
In this work, we propose an extensible framework designed to enhance simulation performance by optimizing both data locality and computational efficiency, thereby addressing these challenges.
This framework is seamlessly integrated with an optimizer that restructures quantum circuits and a simulator that adjusts execution strategies for various quantum operations.
For the newly developed components, \emph{merge booster} and \emph{diagonal detector}, the underlying algorithms are inspired by the principles of quantum entanglement and gate fusion, as well as by the limitations identified in existing third-party simulation libraries.
The experiments were conducted on eight DGX-H100 workstations, each equipped with eight NVIDIA H100 GPUs, employing both gate-level and circuit-level benchmarks. The results indicate a speedup of up to 160 times for circuit-level benchmarks and an acceleration of up to 34 times for diagonal-heavy gate-level benchmarks compared to existing simulators.
The proposed methodologies are anticipated to deliver more robust and faster quantum circuit simulations, thereby fostering the advancement of novel quantum algorithms.
\end{abstract}

\maketitle

\section{Introduction}
\label{sec:introduction}
Quantum computing is an emerging technology exploiting quantum mechanics phenomena to address complex problems that pose challenges to classical computers. Different from classical computing, which represents a `0' or a `1' state with a \emph{bit}, a quantum bit (also known as \emph{qubit}) in quantum computing can represent both states simultaneously. This capability allows for the processing of numerous possibilities all at once. A quantum circuit is a sequence of quantum gates, each operating on qubits, and represents a quantum algorithm designed for a specific computation. By mapping target problems to quantum circuit programs, quantum computers hold the potential to revolutionize various fields, such as cryptography, drug discovery, and combinatorial optimization problems.

Quantum circuit simulation mimics the behaviors of quantum systems with classical computers. It is a crucial way to facilitate the development of quantum algorithms, given the prohibitive expense and noise issue (Noisy Intermediate Scale Quantum; NISQ~\cite{NISQ_IBM}) that restricts the use of physical quantum computers.
Quantum circuit simulation can be categorized by its internal representations for quantum states and simulation algorithms. In particular, full-state simulation and amplitude sampling methods are popular choices for building quantum circuit simulators. In this work, we prioritize full-state simulation for its accurate final probability distribution, suitability for debugging intermediate results, and efficiency in handling complex problems with deep quantum circuits, in contrast to amplitude sampling methods.

The major challenge of full-state quantum circuit simulation is the exponential growth of memory space and computations. In particular, simulating an $N$-qubit quantum circuit requires the memory space to keep $2^\textit{N}$ quantum states. Mathematical operations are required to compute the effects of the quantum gates in the circuit, and they need $2^\textit{N}$ state updates for each simulated gate.
A widely adopted strategy to address this challenge involves the use of clustered computing environments~\cite{QuEST, Doi_2020, imamura2022mpiqulacs, cpu_gpu_communication}, which distribute both quantum states and computations across multiple nodes.
Research efforts have also investigated leveraging storage capacity to extend the number of simulatable qubits~\cite{Hsu31122024}. 
To utilize the computational power of multicore CPUs and GPUs, several studies have incorporated multithreading techniques~\cite{cuQuantum, Cirq, qsim, qHiPSTER, Qiskit, Qulas, QuEST, quantum_plus2}.

Furthermore, various techniques for optimizing quantum circuits have been developed to transform input circuits into more efficient configurations, thereby enhancing overall simulation performance.
Gate fusion, guided by cost functions, consolidates multiple gates into a single equivalent operation, significantly reducing computational overhead~\cite{Aer_fusion}.
Qubit reordering involves remapping qubit indices to minimize memory movement, a particularly effective strategy in multi-device simulations~\cite{Doi_2020, cpu_gpu_communication, cuQuantum}.
To improve data locality in both single-node and multi-node environments, quantum gates of an input circuit can be partitioned into \emph{gate blocks}, accompanied by enhanced data access schemes, providing an effective solution~\cite{HyQuas, uniq, atlas, Queen}.
Detailed knowledge on these optimizations and limitations of prior approaches is described in Sections~\ref{sec:bk-related_work} and \ref{sec:motivation}. 

To extend previous approaches with novel optimization techniques, we revisit quantum circuit partitioning and gate fusion methods, resulting in a systematic framework designed to enable efficient simulation.
The proposed methodologies are anchored in two primary components: \emph{\textbf{swarm optimization}} and \emph{\textbf{adaptive simulation}}.
The swarm optimization component integrates most of the leading optimization strategies for quantum circuits, while further incorporating innovative boosters applicable to a wide range of scenarios and an overhead-free gate fusion technique tailored for diagonal gates.
The adaptive simulation component executes the optimized circuits generated by the swarm optimization, employing the most suitable simulation strategy at the most appropriate execution point.
Consistent with the principle of extensibility, both components are designed to support additional customized optimization approaches.

To validate our approach, the compact and lightweight quantum circuit simulator~\cite{Queen} is extended with the proposed methodologies and an MPI-based communication across the entire cluster.
The experimental evaluation was conducted on a high-performance computing (HPC) cluster equipped with eight DGX-H100 workstations, demonstrating substantial performance improvements over leading simulators, including QuEST~\cite{QuEST}, AerSimulator~\cite{Aer_sim}, cuQuantum~\cite{cuQuantum}, and HyQuas~\cite{HyQuas}.
The experiments encompass both gate-level and circuit-level benchmarks, with a detailed exploration of specific optimizations.
Notably, our improved simulator achieves a 34× speedup over the mainstream simulators in executing the 38-qubit quantum Fourier transform program.
The contributions of this work are summarized as follows.

% \ccw{Experiments conducted on a DGX-H100 server demonstrate significant performance improvements compared to \ccw{leading} simulators, including QuEST~\cite{QuEST}, AerSimulator~\cite{Aer_sim}, cuQuantum~\cite{cuQuantum}, and HyQuas~\cite{HyQuas}. The experiment encompasses both gate-level and circuit-level benchmarks, with a detailed exploration of specific optimizations. Notably, our improved simulator achieves a 5.5× speedup over the mainstream simulators employing a hybrid approach in executing the Quantum Fourier Transform program.}
%Notably, our enhanced simulator achieves a 5.5× speedup over HyQuas~\cite{HyQuas} when running the QFT program.

\begin{itemize}[topsep=2pt,itemsep=0.3em,parsep=0pt]
    \item A systematic and extensible simulation framework, incorporating circuit optimization and simulation, is proposed to expedite full-state quantum circuit simulations.

    \item An adaptive simulator not only supports the most efficient implementation of quantum gates but also selects the most appropriate computation strategy based on the optimized circuit. In particular, a proprietary implementation of fused quantum gates is developed to overcome the practical limitations of existing libraries, as presented in Section~\ref{sec:adaptive_sim}.

    \item A swarm optimizer incorporates the most effective techniques for quantum circuits. Given that this procedure can be categorized as an NP-hard problem~\cite{atlas}, the complexity of all employed techniques is strictly constrained to polynomial-time to prevent transferring execution time from the simulation to the optimization stage, as described in Section~\ref{sec:swarm_opt}.

    \item A novel algorithm \emph{merge booster} is developed to take advantage of the inter-gate block entanglement-free property. It improves computational efficiency while minimizing memory overhead, as will be detailed in Section~\ref{sec:mb}. 

    \item A novel algorithm \emph{diagonal detector} is developed to consider the \emph{commuting gates} property while performing gate fusion. It significantly improves performance when simulating quantum circuits with a high proportion of diagonal gates, as described in Section~\ref{sec:dd}.

    \item A series of benchmarks is employed to evaluate the efficiency of the proposed methodologies. The simulator~\cite{Queen} is extended with our new methodologies and executed on a HPC cluster comprising a total of 64 NVIDIA H100 GPUs. The enhanced simulator consistently delivers superior performance, particularly in large-scale simulations.
    
    % \item A gate-level benchmark and circuit-level benchmarks are used to evaluate the efficiency of our proposed methodologies. We extend the simulator~\cite{Queen} with the proposed methodologies and conduct the experiments on a DGX-H100 workstation with eight NVIDIA H100 GPUs. The enhanced simulator consistently delivers the best performance across all evaluated benchmarks.
\end{itemize}
\section{Background and Related work}\label{sec:bk-related_work}
This section introduces the necessary background information and related works on quantum circuit simulations, with particular emphasis on full-state simulation.
The quantum bit and state representations are described in Section~\ref{sec:statevec}. 
Mathematical formulations of quantum gate operations are provided in Section~\ref{sec:gate}.
A qubit reordering technique for improving data locality is presented in Section~\ref{sec:qubit_swapping}.
Subsequently, the data access schemes employed in quantum circuit simulation are detailed in Section~\ref{sec:schemes}.
Ultimately, existing efforts are reviewed in Section~\ref{sec:related_work}.

\subsection{Quantum Bit and State}
\label{sec:statevec}
Quantum bits (or \emph{qubits}) are basic units of maintaining information for quantum computing. The state of a qubit can be characterized by a column vector $[\alpha \ \ \beta]^T = \alpha \ket{0} + \beta \ket{1}$,
where $\alpha$ and $\beta$ are complex-valued numbers and equate to $1$ in $L_2$ norm. This is also referred to as a quantum state vector in this work. 
The basis states $\ket{0}$ and $\ket{1}$, corresponding to the column vectors $[1 \ \ 0]^T$ and $[0 \ \ 1]^T$
in Dirac notation, form the foundation for qubit state representation. 

In the full-state quantum circuit simulation, each state vector is represented as two 64-bit floating-point numbers, known as a \emph{complex amplitude}.
Extending this concept to an $N$-qubit system, the quantum state can be formulated as following notation.
This representation encapsulates the superposition of all possible basis states in the $N$-qubit system. 

\begin{equation}
\ket{\phi} = \sum_{i \in [0, 2^N)} a_i \ket{i}
\label{eq:state}
\end{equation}

\subsection{Quantum Gate Operation}
\label{sec:gate}
A gate operation in quantum computing performs a specific mathematical transformation on one or more qubits within a quantum circuit. A sequence of quantum gate operations can be represented as a series of matrix operations. 

Considering a single-qubit gate operation $U$ applied on the $j$-th qubit of the state vector, the transformation can be denoted as $U_j = I^{\otimes N-j-1} \otimes U \otimes I^{\otimes j}$, where $U_j$ is a $2^N{\times}2^N$ matrix multiplication on the state vector. The matrix operation on a quantum state $\ket{\psi}$ can be formulated as follows, where $\alpha$ and $\alpha'$ denote the amplitude before and after the operation.

\begin{equation}
\begin{bmatrix}
\alpha'_{b_{N-1}b_{N-2} ... 0_j ... b_{0}}\\ 
\alpha'_{b_{N-1}b_{N-2} ... 1_j ... b_{0}}
\end{bmatrix} = 
U
\begin{bmatrix}
\alpha_{b_{N-1}b_{N-2} ... 0_j ... b_{0}}\\ 
\alpha_{b_{N-1}b_{N-2} ... 1_j ... b_{0}}
\end{bmatrix}
\end{equation}

% 2-qubit gate
A two-qubit $4\times4$ unitary gate $V$ performing the matrix multiplication on the $j$-th and $k$-th qubits can be written as follows, where $j$ is strictly greater than $k$.

\begin{equation}
\begin{bmatrix}
\alpha'_{b_{N-1}b_{N-2}...0_j...0_k...b_{0}}\\ 
\alpha'_{b_{N-1}b_{N-2}...0_j...1_k...b_{0}}\\
\alpha'_{b_{N-1}b_{N-2}...1_j...0_k...b_{0}}\\
\alpha'_{b_{N-1}b_{N-2}...1_j...1_k...b_{0}}
\end{bmatrix} = 
V
\begin{bmatrix}
\alpha_{b_{N-1}b_{N-2}...0_j...0_k...b_{0}}\\ 
\alpha_{b_{N-1}b_{N-2}...0_j...1_k...b_{0}}\\
\alpha_{b_{N-1}b_{N-2}...1_j...0_k...b_{0}}\\
\alpha_{b_{N-1}b_{N-2}...1_j...1_k...b_{0}}
\end{bmatrix}
\end{equation}

In a multithreaded simulation, each thread is typically designated to execute a matrix operation.
As the number of target and control qubits for a given gate increases, the number of threads that can be utilized decreases, and each thread manages an exponentially larger workload.
To alleviate this challenge for certain specialized gates, finer-grained optimization can be implemented~\cite{QuEST}.

\subsection{Qubit Reordering and Qubit Permutation}
\label{sec:qubit_swapping}
In an $\mathit{N}$-qubit simulation, all $2^\mathit{N}$ states are constructed by combining the individual states of each qubit through a tensor product.
When the qubit order is reordered, it solely impacts the arrangement of qubits within a state vector.
Equation~\ref{eq:swap_gate} expresses the operation of a swap gate for the $i$-th qubit that exchanges the quantum states of the $i$-th and $j$-th qubits.
Based on the results, the state of the $i$-th qubit in state vector $a$ is identical to the $j$-th qubit in statevector $a^{\prime}$.

\begin{equation}
\begin{bmatrix}
a^{\prime}_{*...*1_{i}*...*0_{j}*...*} \\
a^{\prime}_{*...*0_{i}*...*1_{j}*...*}
\end{bmatrix}
\mapsto
\begin{bmatrix}
a^{}_{*...*0_{i}*...*1_{j}*...*} \\
a^{}_{*...*1_{i}*...*0_{j}*...*}
\end{bmatrix}
\label{eq:swap_gate}
\end{equation}

This technique can directly influence the permutation within a quantum circuit. For instance, consider a 5-qubit system with the arrangement $(q_4q_3q_2q_1q_0)$.
To swap the positions of $q_3$ and $q_2$, it is essential to exchange the pairs of amplitudes corresponding to $\ket{* 0_3 1_2 * *}$ and $\ket{* 1_3 0_2 * *}$ in the mathematical representation.
In quantum hardware, this is typically achieved by applying a SWAP gate between $q_3$ and $q_2$ to meet connectivity requirements and minimize the accumulation of gate errors~\cite{yan2025quantumcircuitsynthesiscompilation, liu2022swapscostcaseoptimizationaware, pinacanelles2025improvingbenchmarkingnisqqubit}.

The intentional integration of supplementary swaps within the circuit highlights the requirement to modify the indexing of subsequent gates.
In state vector-based simulation, while this method does demand additional operations, the qubits for the following gates have already been rearranged to occupy the least significant bits for the classical computer.
This arrangement signifies that the state represented by that qubit has been positioned closer to the computational unit, resulting in substantial improvements~\cite{Doi_2020, cuQuantum}.

% \subsection{Measurement}
% \label{sec:measurement}
% In quantum computing, measurement involves the extraction of classical information from a quantum system. Different from classical measurements, which observe the state of a system, quantum measurements have effects on the measured system. A measurement will collapse a qubit's superposition of states into one of its basis states, altering the state of the qubit.

% When measuring a specific qubit in the quantum state $\ket{\psi}$, the result can be either $0$ or $1$, each with a distinct probability. 
% For instance, a single-qubit system with the measurement operators can be denoted as $M_0 = \ket{0} \bra{0}$ and $M_1 = \ket{1} \bra{1}$. 
% These operators filter out the amplitudes associated with the eigenstates $\ket{0}$ and $\ket{1}$ on a given $1$-qubit state $\ket{\psi} = \alpha \ket{0} + \beta \ket{1}$.
% The probability of observing the outcome $0$ after the measurement operation is denoted as $\bra{\psi} M_0^{\dag} M_0 \ket{\psi}$, given by a scalar of ${|\alpha|}^{2}$.

% The formulation can be extended naturally to an N-qubit system. 
% Let $M_{i}^j = I^{\otimes n-j-1} \otimes M_i \otimes I^{\otimes j}$ be the measurement operator that filters the amplitudes corresponding to the $j$-th qubit being in the state $\ket{i}$.
% The probability $\bra{\psi} {M_{i}^j}^{\dag} M_{i}^j \ket{\psi}$ becomes the likelihood that the $j$-th qubit is in the state $\ket{i}$. 
% This measurement framework allows for the analysis of multi-qubit systems, providing an understanding of the probabilities associated with specific qubit states.

\subsection{Simulation Scheme} \label{sec:schemes}
\begin{figure}[tb!]
\centerline{\includegraphics[width=.99\columnwidth]{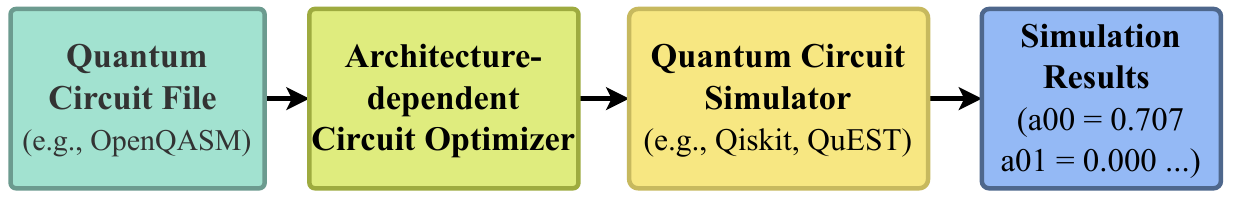}}
\caption{Workflow of quantum circuit simulation.}
\label{fig:workflow}
\end{figure}

\figurename~\ref{fig:workflow} illustrates a typical workflow adopted by modern quantum circuit simulation.
The input consists of a structured file in a specific format (e.g., Quil~\cite{quil} and OpenQASM~\cite{OpenQASM}) to represent a raw quantum circuit.
Subsequently, a quantum circuit optimizer, similar to a quantum compiler, is proficient in performing various quantum circuit optimizations, such as combining sequential quantum gates to reduce circuit depth or overall gate count. The optimizer tends to implement optimizations tailored for the characteristics of a quantum circuit simulator. 
The simulator mimics the behavior of quantum gates specified in the input quantum circuit on a classical computer and generates all amplitudes that comprise the final simulation result.
The following paragraphs present three major schemes for full-state quantum circuit simulation.  
The details of the technical terminology and simulator configurations in this work are provided in Section~\ref{sec:parameters}.

\paragraph{\textbf{Gate-by-Gate Simulation Scheme.}}
In this naive scheme, quantum gates specified in the input quantum circuit are evaluated one by one, and the entire state vector space is traversed and updated after each quantum gate evaluation. The pseudocode of the gate-by-gate simulation scheme is given in Algorithm~\ref{algo:gate_by_gate}.
This scheme is intuitive to implement and is widely adopted by numerous commercial simulators~\cite{Aer_sim, QuEST}.
However, it would incur a significant amount of cache misses when the size of the quantum states exceeds the size of the storage space (e.g., the last-level cache of the processor).
In this case, accessing the quantum states dominates performance, and it would greatly improve performance if the states to be accessed could be cached to avoid accessing lower-level memory frequently.
\begin{algorithm}[tbh!]
    \caption{Gate-by-gate simulation scheme.}
    \label{algo:gate_by_gate}
    \begin{algorithmic}[1]
    \Procedure{gateByGateSim}{\textit{stateVec}, \textit{circuit}}
    \For {\textit{gate} in \textit{circuit}}
        \State \textit{stateVec} $\gets$ operate(\textit{gate}, \textit{stateVec})
    \EndFor
    \EndProcedure
    \end{algorithmic}
\end{algorithm}

\paragraph{\textbf{Block-by-Block Simulation Scheme.}}
The core concept of \emph{block-by-block} simulation can be brought forward in earlier research~\cite{H_ner_2017}.
It is developed to simulate an input quantum circuit in a block-by-block style, thereby alleviating the memory access overhead while updating the quantum states.
Its simulation methodology is specified in Algorithm~\ref{algo:block_by_block}, where the regions subject to optimization and evaluation are collectively referred to as \emph{gate blocks}.

These gate blocks are identified and optimized prior to the simulation. During the simulation, the quantum gates are evaluated in a block-by-block style to achieve data locality.
Upon the evaluation of a gate block, the state vectors of these quantum gates can be accommodated in the higher-level memory of classical processors and processed consecutively, and are designated as \emph{chunks}. 
The technique that transforms an input quantum circuit to facilitate the block-by-block scheme is referred to as \emph{cache blocking} in this work.

\begin{algorithm}[htp!]
    \caption{Block-by-block simulation scheme.}
    \label{algo:block_by_block}
    \begin{algorithmic}[1]
    \Procedure{blockByBlockSim}{\textit{stateVec}, \textit{gateBlocks}}
    \For {\textit{block} in \textit{gateBlocks}}
        \For {\textit{chunk} in \textit{stateVec}}
            \For {\textit{gate} in \textit{block}}
                \State \textit{chunk} $\gets$ operate(\textit{gate}, \textit{chunk})
            \EndFor
        \EndFor
    \EndFor
    \EndProcedure
    \end{algorithmic}
\end{algorithm}

In contrast to the prior gate-by-gate simulation, the caching blocking approach requires an optimizer to accurately determine appropriate gate blocks for consecutive chunk operations in subsequent simulations.
The caching blocking approach can be divided into two classes: the on-demand strategy~\cite{Doi_2020, cpu_gpu_communication} and the strategy aimed at maximizing the size of the gate block~\cite{HyQuas, uniq, Queen, atlas}.
While the former class is straightforward and executes rapidly during the optimization step, the latter allows for more efficient overall simulation, which constitutes the primary focus of this work.

To realize this process, three feasible methods have been proposed to ensure that the required qubits, which are operated by the gates within a gate block, are kept in chunks.
The first method involves the use of multiple fused-swap gates~\cite{Doi_2020, imamura2022mpiqulacs}.
The second method transforms the gate operations to bypass quantum matrix computations, thereby enabling simultaneous swapping for \textit{N} pairs of qubits~\cite{Queen}.
The third method leverages direct indexing to efficiently store data in the cache~\cite{HyQuas}, which is recognized as the most efficient approach, especially for single-machine-node scenarios.

\paragraph{\textbf{Hybrid Simulation Scheme.}}
A \emph{hybrid} simulation scheme is proposed to further accelerate quantum circuit simulations by switching between the two aforementioned schemes, as detailed in Algorithm~\ref{algo:hybrid}. This approach is particularly advantageous when specialized hardware is available for computational acceleration, such as Tensor Cores.
In general, the scheme employs a block-by-block simulation strategy, referred to as \emph{SharedMem} in the original work~\cite{HyQuas}.
However, when a gate acts on a sufficiently large number of qubits, the evaluation of its operations becomes analogous to a general matrix multiplication (GeMM).
In such cases, the hybrid scheme switches to a gate-by-gate simulation mode, termed \emph{BatchMV}, and leverages high-performance matrix multiplication libraries, such as cuBLAS~\cite{cublas} and cuQuantum~\cite{cuQuantum}, to efficiently execute these operations.
It is noteworthy that an additional matrix transpose, e.g., via cuTT~\cite{cutt}, is required to reshape the data layout into a format compatible with cuBLAS\footnote{According to HyQuas~\cite{HyQuas}, the state can then remain in this configuration without requiring reordering back to its original layout.}~\cite{HyQuas}.
\begin{algorithm}[htp!]
\caption{Hybrid simulation scheme.}
\label{algo:hybrid}
\begin{algorithmic}[1]
\Procedure{hybridSim}{\textit{stateVec}, \textit{gateBlocks}}
\For {\textit{block} in \textit{gateBlocks}} 
    \If  {\textit{isFusedGate}} \com{\Comment{Gate-by-gate Scheme (BatchMV)}}
        \State \textit{transVec} $\gets$ transpose(\textit{stateVec})
        \State \textit{stateVec} $\gets$ operateGeMM(\textit{fusedGate}, \textit{transVec})
    \Else \com{\Comment{Block-by-block Scheme (ShareMem)}}
        \State blockByBlockSim(\textit{stateVec}, \textit{gateBlocks})
        % \For {\textit{chunk} in \textit{stateVec}}
        %     \For {\textit{gate} in \textit{block}}
        %         \State \textit{chunk} $\gets$ operate(\textit{gate}, \textit{chunk})
        %     \EndFor
        % \EndFor
    \EndIf
\EndFor
\EndProcedure
\end{algorithmic}
\end{algorithm}

\subsection{Related Work}
\label{sec:related_work}
Extensive research has been conducted on quantum circuit simulations. Various internal data representations, such as state vectors, tensor network states, and density matrices, have been employed by quantum circuit simulators, each exhibiting distinct advantages and drawbacks.
Among these, the state vector method is particularly valued for its noise-free characteristics, attracting significant attention and optimization efforts from both quantum and classical computing researchers.
However, the full potential performance of state vector-based simulation still needs to be realized, consolidated, and integrated.
Hence, the fundamental acceleration techniques and scalability enhancements for both the quantum circuit and simulation, as presented below, have been systematically organized.

\paragraph{\textbf{Parallel Simulation.}} Numerical efforts have been made for parallelizing state vector-based quantum circuit simulations that exploit data-level parallelism. Remarkable examples, such as qHiPSTER~\cite{qHiPSTER}, qsim~\cite{qsim}, and QuEST~\cite{QuEST}, take advantage of CPU multithreading support to boost simulation performance. There are works, such as Cirq~\cite{Cirq} and qHiPSTER~\cite{qHiPSTER}, leveraging the single-instruction-multiple-data (SIMD) support on multicore processors for simulation acceleration. More recently, GPU-based approaches have been used to improve simulation efficiency, with key examples including Aer-Simulator~\cite{Aer_sim}, cuQuantum~\cite{cuQuantum}, HyQuas~\cite{HyQuas}, and Queen~\cite{Queen}.
The performance of cuQuantum and HyQuas is impressive in both cross-node and single-node simulations, respectively. Queen demonstrates both aspects but has yet to explore the potential benefits of third-party libraries.

\paragraph{\textbf{Scalable Simulation.}} Typically, state vector-based quantum circuit simulations rely on primary memory to maintain full quantum states.
To extend the capacity for simulating a larger number of quantum bits, one possible approach is to incorporate memories from different machine nodes.
This allows a larger number of quantum state vectors to be distributed to these nodes. The Message Passing Interface (MPI) is commonly used to facilitate the necessary communication involving data transfers between nodes~\cite{QuEST, cpu_gpu_communication, rdma}.
For efficient GPU-based simulation, NVIDIA Collective Communications Library (NCCL)~\cite{nccl} utilizes NVLink to achieve superior interconnect throughput across multiple GPUs~\cite{HyQuas, atlas, Queen}.
An alternative, cost-effective solution has been suggested, which involves leveraging secondary storage devices, such as SSDs, to support larger-scale quantum circuit simulations. In comparison to NVLink methods, intra-machine disk I/O operations can also enhance the simulation efficiency~\cite {Hsu31122024}.

\paragraph{\textbf{Quantum Circuit Optimization.}}
Additionally, quantum circuit optimizations can be employed to further accelerate simulations.
These optimizations transform input circuits into more streamlined versions before the simulation begins, thereby reducing simulation time through techniques such as gate fusion and qubit reordering.
Gate fusion is a well-established technique that combines consecutive quantum gates into a single generic quantum gate to minimize memory access overhead.
Despite its advantages, fused general gates typically require a cost function to maintain parallel efficiency~\cite{Aer_fusion, HyQuas}.
Qubit reordering modifies the data access pattern of qubits used in quantum algorithms to improve data locality~\cite{cpu_gpu_communication, Doi_2020, cuQuantum, Queen}.
The comprehensive implementation strategies for both intra-node and inter-node qubit exchanges have been thoroughly explored in Queen~\cite{Queen}.
While potentially lowering readability for developers, encoding gate information into a bitmap can confer a substantial advantage in optimizer efficiency~\cite{HyQuas, uniq}.

These diverse optimization scopes have established a robust foundation, demonstrating the effectiveness of full-state simulation when combined with other supportive methodologies.
% To further extend these techniques, a comprehensive framework is proposed that systematically integrates all three key features. Section~\ref{sec:workflow} discusses the limitations of current optimization techniques. Section~\ref{sec:aio_optization_module} introduces the new framework, which comprises two major modules and two core techniques. Subsequently, Section~\ref{sec:simulation_module} presents the performance results obtained on an HPC cluster.
% presents the performance results obtained on an HPC cluster.

% In this work, these considerations are systematically integrated and further extended through additional optimization techniques.

\section{Motivation} \label{sec:motivation}
This section addresses the limitations of existing works, concentrating on three main aspects. 
First, the challenges related to portability and efficiency of the existing library in the hybrid scheme are presented.
Second, the constraints of current gate block search algorithms in the block-by-block scheme are examined.
Third, the shortcomings of existing gate fusion algorithms in the gate-by-gate scheme are addressed.

\paragraph{\textbf{Limitation of the Adoption of Existing Library.}}
State vector-based simulation inherently provides a structured quantum representation, enabling the integration of diverse optimization techniques suited to various scenarios.
A wide spectrum of advanced approaches has been proposed across numerous significant studies, making it impractical for any single framework to encompass them all.
Conversely, the indiscriminate integration of every available technique can result in excessive complexity, ultimately hindering maintainability.
Specifically, the use of third-party libraries~\cite{cutt, cublas} to accelerate the computations for fused gates forces a reversion to the gate-by-gate scheme, as shown in Algorithm~\ref{algo:hybrid}.

These methods not only fail to preserve the cached property but often necessitate the eviction of cached data from shared memory to accommodate input loading.
Furthermore, relying on such libraries significantly amplifies memory consumption and data transfer overhead, potentially leading to unmanageable out-of-memory issues and excessive data transfers during simulations.
In contrast, implementing the \emph{All-in-one simulation}~\cite{Queen} can help mitigate these challenges by reducing unnecessary resource overhead and enhancing the capacity for flexible optimization mechanisms.

\paragraph{\textbf{Limitation of Gate Block Search Algorithms.}}
A widely recognized strategy for accelerating large-scale quantum circuit simulations involves utilizing a block-by-block scheme through gate block search algorithms, which effectively decompose a raw circuit into multiple gate blocks.
These search algorithms are typically designed to maintain polynomial-time complexity~\cite{HyQuas, Queen, Doi_2020}, ensuring that simulations can be conducted within a practical timeframe for online applications.
Regardless of how the circuit is divided into various blocks, full access to the state vector remains essential. 

\begin{figure}[tb!]
\centerline{\includegraphics[width=0.95\columnwidth]{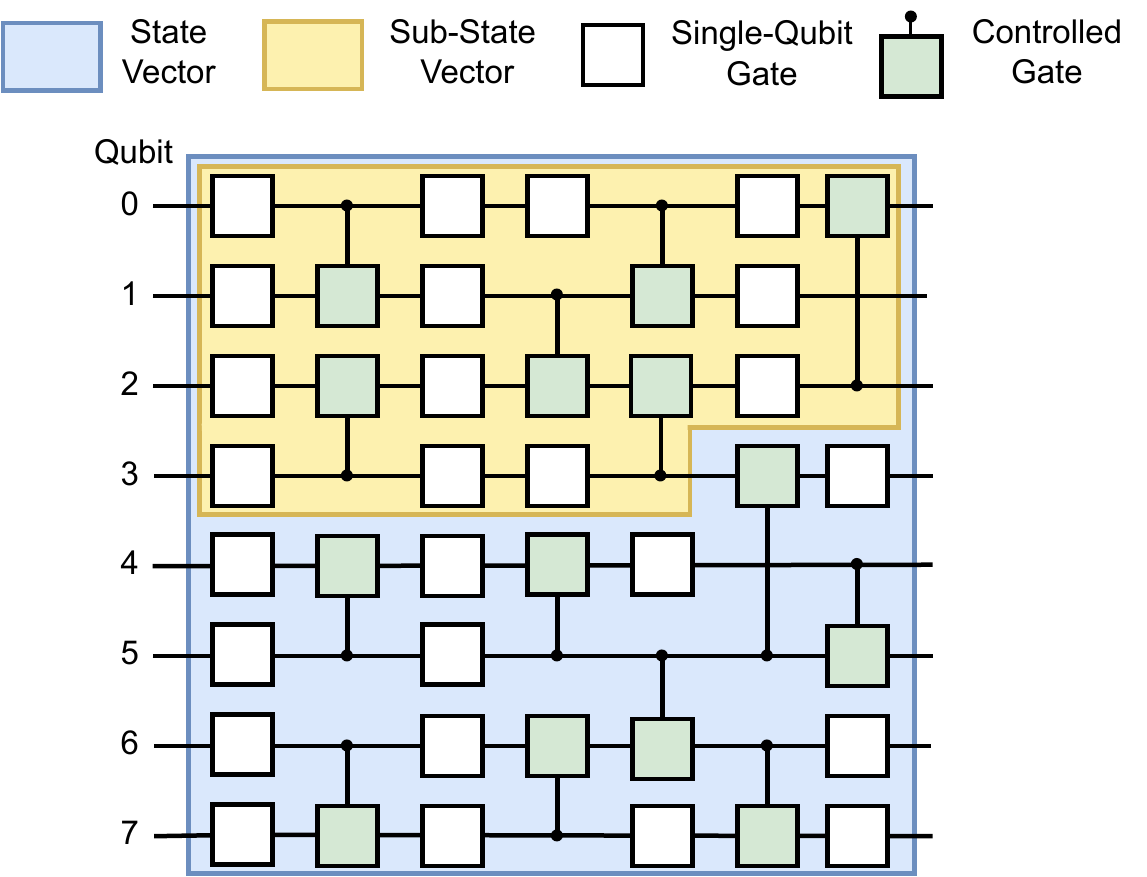}}
\caption{
Memory allocation and quantum gate arrangements for the standard state vector-based simulation. The blue and yellow regions indicate the quantum operations applied to the full-state vector and the sub-state vector, respectively.
}
\label{fig:case1}
\end{figure}

\figureautorefname{~\ref{fig:case1}} illustrates an 8-qubit quantum circuit consisting of 40 gates, including 26 single-qubit gates and 14 controlled gates. Simulating this circuit requires a total of $40 \times 2^{8}$ state vector updates. However, although the highlighted (yellow) gate block operates on only 4 qubits, it still incurs $21 \times 2^{8}$ state vector updates.
This access pattern represents a common limitation of contemporary full-state vector simulators~\cite{QuEST, Aer_sim, HyQuas, cuQuantum, Cirq, uniq, atlas}.
One of our key findings is that we can further reduce the number of state vector updates when the quantum gates in a partitioned quantum circuit are not entangled with the qubits outside of this partitioned circuit. 
This scenario is referred to as an \emph{inter-gate block entanglement-free situation}. 
In the aforementioned example, since there is no entanglement between the qubits within the highlighted region and those external to it, the required number of state updates can be diminished to $19 \times 2^{4}$.
Capitalizing on this characteristic can result in a substantial reduction in memory access overhead; conversely, overlooking it may restrict potential performance enhancements.

The implementation of this optimization entails additional costs, such as the need for extra memory buffers and supplementary tensor product operations. 
That is, when a parallel thread processes a partitioned quantum circuit, an auxiliary buffer becomes necessary for storing intermediate results, alongside a tensor product operation for reshaping the computed data to align with the layout of the original state vector.
If this optimization is not meticulously managed, it may result in degraded computational efficiency and increased memory consumption.
To address these trade-offs comprehensively, this work proposes the \emph{\textbf{merge booster}} to identify suitable gate blocks within an input quantum circuit, striking a balance between memory utilization and computational overhead.

\paragraph{\textbf{Limitation of Gate Fusion Algorithms.}}
In addition to modifying the access pattern, gate fusion is a commonly adopted optimization technique in state vector-based quantum circuit simulations for reducing computational overhead.
It works by combining multiple consecutive quantum gates into a single equivalent matrix operation (e.g., arbitrary unitary gates), thereby minimizing the number of matrix multiplications required to simulate the original consecutive quantum gates.
In particular, diagonal fusion is a specialized form of gate fusion, targeting consecutive diagonal gates, such as the Pauli-Z (Z) gate, the controlled-phase (CP) gate, and the two-qubit ZZ rotation (RZZ) gate.
This optimization is particularly useful for quantum circuits containing a significant number of diagonal gates, such as the standard Quantum Approximate Optimization Algorithm (QAOA)~\cite{Lin_2024, for_qaoa}.

Existing gate fusion algorithms overlook the fusion opportunities arising from the property of \emph{commuting gates} and miss potential performance improvements enabled by these gates. Two quantum gates are said to \emph{commute} if swapping their order of application does not affect the final quantum state.
Leveraging this property, such gates can be efficiently detected and combined whenever feasible.
As illustrated in \figureautorefname{~\ref{fig:diagOpt}}(a), a naive fusion algorithm stops the fusion process upon encountering the RX$_{7}$ gate while scanning for potential fusion opportunities from top-left to bottom-right of a given quantum circuit. However, when the commuting gate property is taken into account, the RZZ$_{8}$ and CP$_{9}$ gates can be repositioned and fused with earlier gates before the RX$_{7}$ gate, as shown in \figureautorefname{~\ref{fig:diagOpt}}(b). 
In this work, we devise a \emph{\textbf{diagonal detector}} algorithm to take advantage of commuting gates to enlarge the quantum gates to be fused and accelerate simulation speed.
\section{Methodology} \label{sec:method}
In this section, we present the fundamental components of our proposed framework. Section~\ref{sec:parameters} provides an overview of the essential parameter configurations required for quantum circuit simulations. In Section~\ref{sec:swarm_opt}, we outline an integrated collection of efficient circuit optimizations. Furthermore, we discuss the design of the simulation module, which is adaptable to various hardware scenarios, in Section~\ref{sec:adaptive_sim}. Section~\ref{sec:mb} explains the merge booster, designed to streamline computations by eliminating unnecessary state updates. Lastly, Section~\ref{sec:dd} introduces the diagonal detector, employed for implementing gate fusion specifically on diagonal gates.

\subsection{Parameter Configurations} \label{sec:parameters}
The quantum circuit simulation requires several configurations and system parameters.
In the simulation framework, the data structure \textit{env} encapsulates the environment, including critical parameters such as \textit{N}, \textit{B}, and \textit{C}, as well as other relevant metrics.

The parameter \textit{N} signifies the total number of qubits within the simulation. The boosting optimization process is impacted by \textit{B}, where \(2^B\) determines the minimum division size. The parameter \textit{C} denotes the size of the sub-states extracted from the \(2^N\) amplitudes, which must be carefully selected to stay within the cache capacity for optimal performance. Additionally, the parameter \textit{R} indicates the total number of ranks within the simulation system. For fusion strategies, the parameters \textit{D} and \textit{F} specify the maximum number of target qubits processed through diagonal and cost-based gate fusion, respectively.

The simulation also relies on \textit{stateVec}, which holds the quantum state composed of $2^N$ amplitudes. The quantum operations applied during the simulation are represented by \textit{gate}, while \textit{G} signifies the total number of quantum gates within the circuit.
The parameter \textit{targs} specifies the target qubits associated with each gate.
The quantum circuit itself, consisting of a sequence of gates, is labeled as \textit{circuit}.
To enhance efficiency, an optimized arrangement of operations, referred to as \textit{gateBlocks (GBs)}, is derived by considering the original circuit, the applied optimizations, and the system environment.

\subsection{Swarm Optimization Module} \label{sec:swarm_opt}
Algorithm~\ref{algo:swarm_opt} illustrates the overall workflow of the swarm optimization module, which is strictly constrained to operate using polynomial-time algorithms and typically accounts for less than one-thousandth of the total execution time. This module is designed to enhance quantum circuit performance by systematically applying various optimization techniques across multiple hierarchical levels, including rank-level, machine-level, and computation-level operations.

The gate block search algorithm (GBSA) builds upon the design of the fundamental simulator.
This algorithm utilizes a unified function to separate circuits into several non-overlapping subcircuits, effectively tackling similar partitioning optimization problems.
To simplify the overall workflow, the proposed GBSA is repeatedly invoked to jointly enable gate fusion and cache blocking techniques.
Additionally, to improve performance across various dimensions, this algorithm can be replaced or extended as noted in previous works~\cite{HyQuas, atlas, SnuQS, Doi_2020}.
To support the new optimization techniques, the merge booster produces entanglement-free blocks with efficiency, avoiding unnecessary quantum state updates.
The diagonal detector consolidates consecutive diagonal gates into a single gate without specific ordering, simultaneously reducing both the number and depth of gates.

\begin{algorithm}[hbt!]
    \caption{Swarm optimization.}
    \label{algo:swarm_opt}
    \begin{algorithmic}[1]
    \Require \textit{N}, \textit{R}, \textit{C}, \textit{F}, \textit{B}, \textit{circuit}
    \State \textit{devGBs} $\gets$ GBSA(\textit{N}, \textit{N}-\textit{R}, \textit{circuit}, 0) \com{\Comment{Rank-level}}
    \For{\textit{subCircuit} in \textit{devGBs.size()}} \com{\Comment{Machine-level}}
        % \If {i == 0} 
        \State \textit{GBs} $\gets$ Booster(\textit{N}, \textit{C}, \textit{B}, \textit{F}, \textit{subCircuit}) 
        % \com{\Comment{Boosting-block}}
        % \EndIf
        \State \textit{GBs} $\gets$ DiagonalDetector(\textit{N}, \textit{D}, \textit{subCircuit}) 
        % \com{\Comment{Diagonal-block}}
    
        \State \textit{GBs} $\gets$ GBSA(\textit{N}, \textit{C}, \textit{subCircuit}, 0) 
        % \com{\Comment{General-block}}

        % \If{\textit{isFusion}}
        \For{\textit{GB} in \textit{GBs}} \com{\Comment{Computation-level}}
          \State \textit{fusedGBs} $\gets$ GBSA(\textit{N}, \textit{F}, \textit{GB}, 1)
        \EndFor
        \State \textit{GBs} $\gets$ \textit{fusedGBs}
      % \EndIf
    \EndFor
    \end{algorithmic}
\end{algorithm}
To summarize, the swarm-inspired optimization approach systematically applies circuit transformations, enhancing simulation efficiency and preserving flexibility for future extensions.

\subsection{Adaptive Simulation Module} \label{sec:adaptive_sim}
The adaptive simulation scheme judiciously selects the most suitable simulation strategy for efficient quantum circuit simulations, as demonstrated in Algorithm~\ref{algo:flow}. This module takes as input the quantum state vector \textit{stateVec} and an optimized sequence of \textit{gateBlocks}.
Each \textit{block} within \textit{gateBlocks} is processed based on one of three primary strategies: tensor product computation, diagonal fused gate simulation, or general gate simulation.

When the tensor product condition is satisfied, the simulator computes the relevant sub-vectors, \textit{subVec1} and \textit{subVec2}, and updates \textit{stateVec} accordingly.
In the case of fused diagonal gates, the simulator applies the fused diagonal operation directly to \textit{stateVec}, thereby circumventing the overhead associated with general-purpose matrix operations.
For conditions requiring the rearrangement of the state vector, the task is delegated to the qubit reordering function.
In instances appropriate for an all-in-one simulation strategy, \textit{stateVec} undergoes processing on a chunk-wise basis.
This approach optimizes data locality and minimizes unnecessary computations, regardless of whether the gates are fused.

For all other promising scenarios, this module provides opportunities for further optimizations through an extensible function region, thereby facilitating future enhancements.

\begin{algorithm}[htp!]
\caption{Adaptive simulation scheme.}
\label{algo:flow}
\begin{algorithmic}[1]
\Procedure{adaptiveSim}{\textit{stateVec}, \textit{gateBlocks}}
\For {\textit{block} in \textit{gateBlocks}}
    \If {isTensorProduct} \com{\Comment{Tensor Product}}
        \State \textit{stateVec} $\gets$ tensorProduct(\textit{subVec1}, \textit{subVec2})
    \ElsIf  {isFusedDiag} \com{\Comment{Diagonal Simulation}}
        \State \textit{stateVec} $\gets$ operate(\textit{fusedDiagGate}, \textit{stateVec})
    % \ElsIf  {isGeMM} \com{\Comment{GeMM Gate Simulation}}
    %     \State \textit{transVec} $\gets$ \texttt{transpose}(\textit{stateVec})
    %     \State \textit{stateVec} $\gets$ \texttt{operate}(\textit{fusedGeMM}, \textit{transVec})
    \ElsIf {isQubitReorder} \com{\Comment{Qubit Reordering}}
        \State \textit{stateVec} $\gets$ reorder(\textit{qubits}, \textit{stateVec})
    \ElsIf {isBlock} \com{\Comment{All-in-one Simulation}}
        \For {\textit{chunk} in \textit{stateVec}}
            \If{isFusedGate}
                \State \textit{chunk} $\gets$ operate(\textit{fusedGate}, \textit{chunk})
            \Else
                \State \textit{chunk} $\gets$ operate(\textit{gate}, \textit{chunk})
            \EndIf
        \EndFor
    \Else \com{\Comment{Applying further extensible optimization}}
        \State doExtensibleOptimization(\textit{stateVec})
    \EndIf
\EndFor
\EndProcedure
\end{algorithmic}
\end{algorithm}

\subsection{Merge Booster} \label{sec:mb}
In Section~\ref{sec:motivation}, the idea of reducing the number of quantum state updates for the quantum gates without inter-gate block entanglement incurs additional memory and computation overheads. The \textbf{\emph{merge booster}} algorithm is devised to balance the simulation efficiency and the incurred overhead. 
A larger gate block can improve simulation efficiency by reducing the number of extra computations required to merge the sub-state vector, but at the cost of allocating a larger memory buffer to store temporary quantum states.
Conversely, a smaller gate block demands less memory but incurs a higher computational cost for merging sub-state vectors back into the original state vector.
This algorithm aims to improve simulation time, memory consumption, and computational overhead.

\begin{figure}[tb!]
\centerline{\includegraphics[width=1\columnwidth]{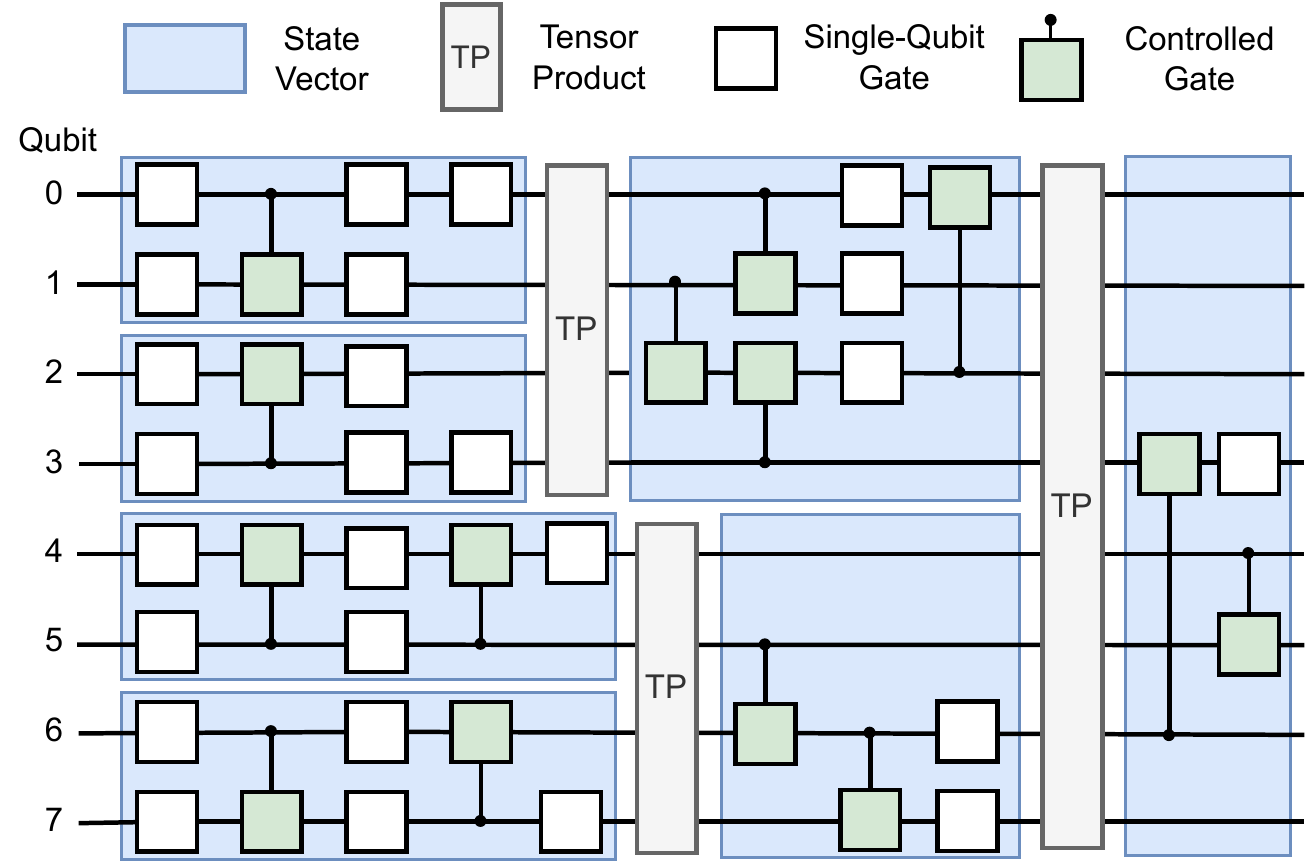}}
\caption{Memory allocation and quantum gate arrangements for the proposed state vector-based simulation. Blue region represents the quantum operation applied to the state vector, and the grey block indicates the tensor product operation.}
\label{fig:case2_tp}
\end{figure}

\paragraph{\textbf{An Illustrative Example.}} 
Given the raw quantum circuit in \figureautorefname{~\ref{fig:case1}}, \figureautorefname{~\ref{fig:case2_tp}} provides a clear illustration of the circuit optimized by the merge booster. 
With 40 gates in the 8-qubit circuit, a naive approach requires $40\times 2^{8}$ accesses to update the $2^{8}$ state vector for each gate operation.
In contrast, the merge booster can reduce the access count to 1,336. The reduced access count is calculated based on quantum gates in $n$-qubit gate blocks: $(6+6+7+7)\times 2^{2}+(7+4)\times 2^{4}+3\times 2^{8}$ for state vector, with an additional $2\times 2^{4}+2^{8}$ for tensor product overheads. Based on this empirical analysis, it can result in a 7-fold improvement in the number of state updates.
Furthermore, the benefits become even more pronounced as the number of simulated qubits increases, with cache blocking and gate fusion seamlessly integrated into the process.
Overall, the merge booster achieves the time complexity of $\mathcal{O}(\textit{N}log(\textit{N})+\textit{G}+\textit{C})$, where \textit{N} represents the number of qubits, \textit{G} denotes the total number of gates, and \textit{C} is the overhead introduced when fusion and blocking techniques are enabled. 
The space complexity is $\mathcal{O}$(\textit{N}), driven by a queue and a one-dimensional vector.

\paragraph{\textbf{Algorithm Design.}} 
The core idea can be considered as an extension of the classical merge sort algorithm, primarily divided into two components: division and merging, as depicted in Algorithm~\ref{algo:booster}. The divide component in Algorithm~\ref{algo:divider} involves reducing the maximum memory usage by powers of two, which can ensure minimal additional memory consumption.
For a 32-qubit simulation, only two 2$^{16}$ sub-state vectors are required to handle the most significant memory block.
Besides, based on the microbenchmarking results for quantum gates, the reduction in qubits no longer yields exponential gains once the state vector can be cached.
Therefore, the divide component provides a mechanism for controlling the minimum size of gate blocks.

\begin{algorithm}[htp!]
\caption{Divide component of Merge Booster.}
\label{algo:divider}
\begin{algorithmic}[1]
\Procedure{\textsc{Divider}}{\textit{N}, \textit{divSize}, \textit{que}}
\If {\textit{N} $\leq$ \textit{divSize}}
    \State que.push(\textit{N})
    \State \textbf{return}
\EndIf
\State \textsc{Divider}(\textit{N} \texttt{>{}>} 1, \textit{divSize}, \textit{que}) \com{\Comment{Halve the queue size}}
% \If {\textit{N} \texttt{\&} 1}
%     \State \textit{N} $\gets$ \textit{N} + 1
% \EndIf
\State \textit{N} $\gets$ (\textit{N} \texttt{\&} 1) ? \textit{N} + 1 : \textit{N}
\State \textsc{Divider}(\textit{N} \texttt{>{}>} 1, \textit{divSize}, \textit{que})
\EndProcedure
\end{algorithmic}
\end{algorithm}
\begin{algorithm}[htp!]
\caption{Pseudocode of the Merge Booster.}
\label{algo:booster}
\begin{algorithmic}[1]
\Procedure{\textsc{MergeBooster}}{\textit{circuit}, \textit{gateBlocks}}
\State que $\gets$ queue()
\State \textsc{Divider}(\textit{N}, (\textit{N} + \textit{B} - 1) / \textit{B}, que)
\While {que.size() \texttt{>} 1} \com{\Comment{Start merge component}}
    \State \textit{accIdx} $\gets$ 1
    \State \textit{mergedVec} $\gets$ []
    \State \textit{mergedSum} $\gets$ 0
    
    \For {\textit{i} in \textit{que}.size()}
        \State \textit{gBlock} $\gets$ []\label{algo:booster:prepare_begin} \com{\Comment{1. Prepare data}}
        \State \textit{boostSet} $\gets$ set()
        \For {\textit{j} in range(\textit{accIdx}, \textit{accIdx} + \textit{que}.front())}
            \State \textit{boostSet}.insert(\textit{j})
        \EndFor \label{algo:booster:prepare_end}

        \com{\Comment{2. Generate the gate blocks}}
        \State \textit{gBlock} $\gets$ genGateBlock(\textit{boostSet}, \textit{circuit})\label{algo:booster:gen_begin}
        % \If {\textit{env.isDiagFusion}}
        \State \textit{gBlock} $\gets$ doFusion(\textit{gBlock})
        % \EndIf
        % \If {\textit{boostSet}.size() \texttt{>} \textit{env.C}}
        \State \textit{gateBlocks} $\gets$ genGateBlock(gBlock)\label{algo:booster:gen_end}
        % \Else
        % \State \textit{gateBlocks}.push\_back(\textit{gBlock})
        % \EndIf\label{algo:booster:gen_end}

        \com{\Comment{3. Merge required data}}
        \State \textit{mergedSum} $\gets$ \textit{mergedSum} + \textit{que}.front()\label{algo:booster:merge_begin}
        \If {i \texttt{\&} 1}
            \State \textit{mergedVec}.push(\textit{mergedSum})
            \State \textit{mergedSum} $\gets$ 0
        \EndIf
        \State \textit{accIdx} $\gets$ \textit{accIdx} + \textit{que}.front()
        \State \textit{que}.pop()
    \EndFor\label{algo:booster:merge_end}
    
    \com{\Comment{4. Update merged states and gate block}}
    \For {\textit{i} in \textit{mergedVec}.size()}\label{algo:booster:update_begin}
        \State \textit{que}.push(\textit{mergedVec}[i])
        \State \textit{qubits} $\gets$ getAllElements(\textit{mergedVec}, {i})
        \State \textit{gateBlocks}.insert(\textit{TensorProduct(\textit{qubits})})
    \EndFor\label{algo:booster:update_end}

\EndWhile
\EndProcedure
\end{algorithmic}
\end{algorithm}

The merge component primarily relies on the collaboration between a queue and an array through their member functions for assigning specific indices to gate blocks.
Moreover, it can be divided into four distinct steps and is fully compatible with both gate fusion and cache blocking, unlike the hybrid strategy~\cite{HyQuas}.

\begin{enumerate}
    \item \emph{Preparation Step}: Extract the required qubits from the queue and place the corresponding data into the set (Lines \ref{algo:booster:prepare_begin}-\ref{algo:booster:prepare_end}). This step ensures that all necessary qubits are available for subsequent processing.

    \item \emph{Generation Step}: Generate the gate block within the \textit{circuit} by identifying gates that involve qubits from the set.
    When the number of required qubits exceeds the available cache size or gate fusion optimization is feasible, the fine-grained gate blocks are generated to optimize both memory usage and computational efficiency (Lines \ref{algo:booster:gen_begin}-\ref{algo:booster:gen_end}).
    
    \item \emph{Merge Step}: Merge the data pairwise, compute the subsequent round of required qubits, and remove the corresponding elements from the queue (Lines \ref{algo:booster:merge_begin}-\ref{algo:booster:merge_end}).

    \item \emph{Update Step}: Update the queue with the newly required qubits, and insert the additional tensor product operations to the \textit{gateblock} (Lines \ref{algo:booster:update_begin}-\ref{algo:booster:update_end}).
\end{enumerate}

\begin{algorithm}[htp!]
\caption{Pseudocode of Diagonal Detector.}
\label{algo:diagonalDetector}
\begin{algorithmic}[1]
\Procedure{\textsc{diagonalDetector}}{\textit{env}, \textit{gateBlocks}}
    \State \textit{resList} $\gets$ [] \com{\Comment{A list of processed gates}}
    \State \textit{it} $\gets$ \textit{gateBlocks.begin}()
    
    \While {\textit{it} != \textit{gateBlocks.end}()}
        \If {!isDiagonal(\textit{it})} \label{line:not_skip_begin} \com{\Comment{Skip other types}}
            \State \textit{resList.push\_back}(*\textit{it})
            \State ++\textit{it}
            \State \textbf{continue} \label{line:not_skip_end} 
        \EndIf
        
        % --- Start diagonal detection ---
        \State \textit{diagList} $\gets$ [*\textit{it}] \com{\Comment{A list of diagonal gates}}
        \State \textit{uList} $\gets$ [] \com{\Comment{A list of unprocessed gates}}
        \State \textit{stopTable} $\gets$ zeros(\textit{env.N})
        \State \textit{depSet} $\gets$ set(\textit{it}-\texttt{>}targs) 
        \State \textit{diagSize} $\gets$ 1
        \State \textit{fIt} $\gets$ next(\textit{it}) 
        \While {\textit{fIt} != \textit{gateBlocks.end}()} \label{line:fIt_begin}
            \State \textit{intSet} $\gets$ intersect(\textit{fIt}-\texttt{>}targs, \textit{depSet})
            
            \If {!isDiagonal(\textit{fIt})} \label{line:not_diag_begin} \com{\Comment{Not a diagonal gate}}
                \If {!\textit{intSet.empty}()}
                    \State \textit{stopTable} $\gets$ update(\textit{stopTable}, \textit{fIt})
                    \State \textit{uList.push\_back}(*\textit{fIt})
                \Else
                    \State \textit{resList.push\_back}(*\textit{fIt})
                \EndIf \label{line:not_diag_end} 
            \Else \com{\Comment{A diagonal gate is detected}}
                \State \textit{depSet.insert}(\textit{fIt}-\texttt{>}targs) 
                \If {checkStop(\textit{stopTable}, \textit{fIt})} \label{line:stop_test_begin} 
                    \State \textbf{break}
                \EndIf \label{line:stop_test_end}
                \State \textit{diagList.push\_back}(*\textit{fIt}) \label{line:connect_begin}
                \State ++\textit{diagSize} \label{line:connect_end}
            \EndIf
            \State ++\textit{fIt}
        \EndWhile \label{line:fIt_end}
        
        \If {\textit{diagSize} \texttt{>} 1} \label{line:do_diag_begin} \com{\Comment{Do diagonal fusion}}
            \State \textit{fusedGate} $\gets$ doDiagonalFusion(\textit{diagList})
            \State \textit{resList.push\_back}(\textit{fusedGate})
        \EndIf \label{line:do_diag_end}
        
        \State \textit{resList.extend}(\textit{uList})
        \State \textit{it} = \textit{fIt}
    \EndWhile
    
    \State \textit{gateBlocks} $\gets$ \textit{resList}
\EndProcedure
\end{algorithmic}
\end{algorithm}

\subsection{Diagonal Detector} \label{sec:dd}
Diagonal gates, such as Z, CP, and RZZ gates, possess unique properties that can be exploited to optimize quantum circuit simulations.
Unlike general quantum gates, diagonal gates only alter the phase of quantum states without affecting probability amplitudes, which makes them ideal candidates for specialized fusion techniques. 
To efficiently identify and fuse the largest gate blocks composed of diagonal gates, we introduce the \textbf{\emph{diagonal detector}} algorithm, which restructures quantum circuits using a linked-list-based approach that captures the dependencies of target and control qubits.

The matrix operation of an \textit{N}-qubit diagonal gate on the \textit{N}-qubit state vector is illustrated below. In a multithreaded execution environment, thread$_{i}$ is tasked only with performing complex multiplication between $\lambda_i$ and $\alpha_i$ of the following equation.

\[
|\psi'\rangle = D_{\textit{N}}|\psi\rangle = 
\begin{bmatrix}
\lambda_1 & 0 & 0 & \cdots & 0 \\
0 & \lambda_2 & 0 & \cdots & 0 \\
% 0 & 0 & \lambda_3 & \cdots & 0 \\
\vdots & \vdots & \vdots & \ddots & \vdots \\
0 & 0 & 0 & \cdots & \lambda_{2^N}
\end{bmatrix}
\begin{bmatrix}
\alpha_1 
\\ \alpha_2 
% \\ \alpha_3 
\\ \vdots 
\\ \alpha_{2^N}
\end{bmatrix}
\]

\begin{figure*}[tb!]
\centerline{\includegraphics[width=1.96\columnwidth]{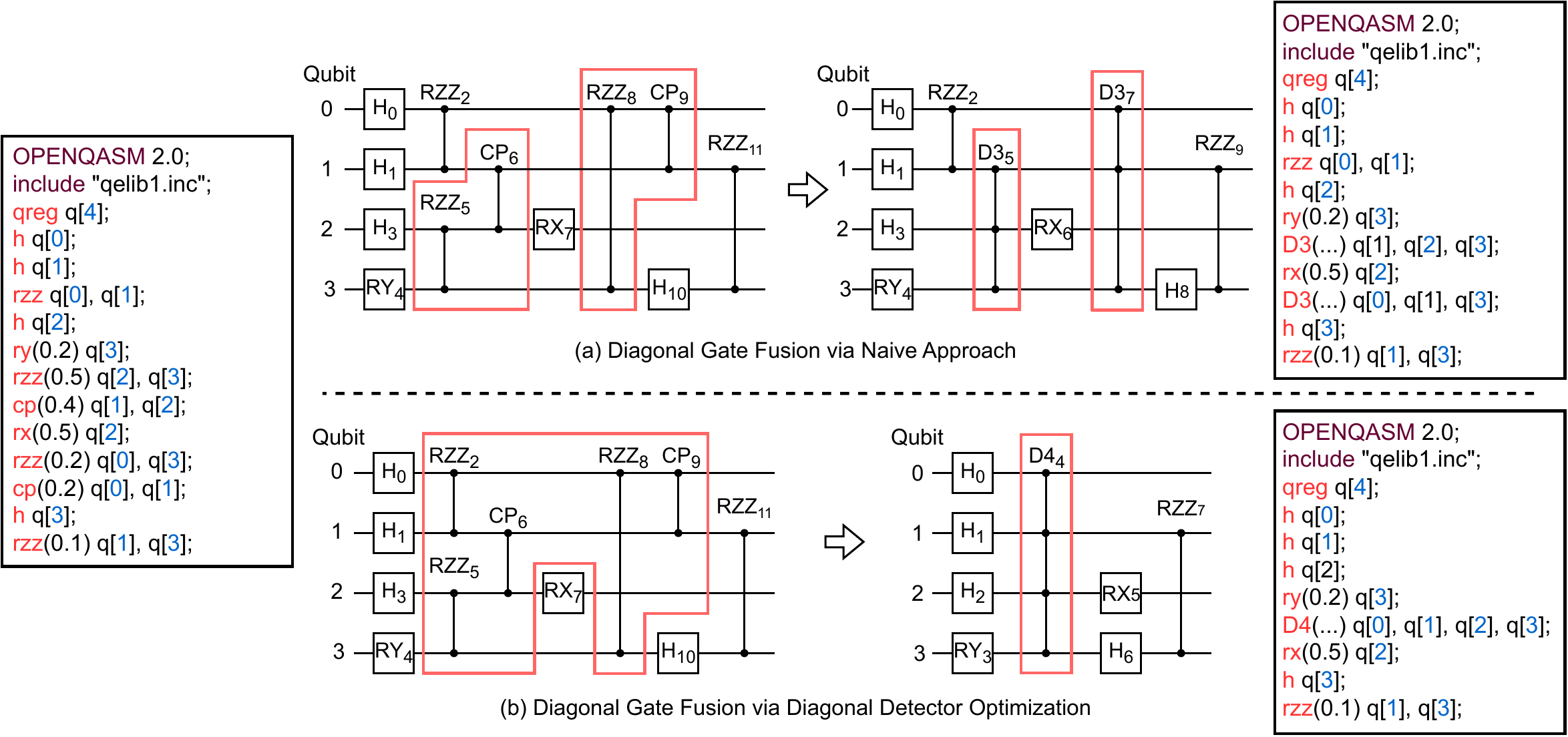}}
    \caption{
        Different ways to enable diagonal gate fusion: (a) a naive approach, (b) the proposed diagonal detector optimization. The left side represents the circuit before optimization, while the right side shows the optimized circuit. The \textit{D} represents a diagonal gate, and the subscript in the gate notation indicates the execution order.
    }
\label{fig:diagOpt}
\end{figure*}

\paragraph{\textbf{An Illustrative Example.}} 
The diagonal detector maximizes fusion opportunities without violating circuit constraints by analyzing gate dependencies and applying set intersections. As shown in Figure 4, our method significantly improves upon conventional approaches, allowing gates like RZZ$_{2}$, RZZ$_{5}$, CP$_{6}$, RZZ$_{8}$, and CP$_{9}$ to be merged efficiently into a single operation D4$_{4}$. On the contrary, a conventional approach merges RZZ$_{5}$ and CP$_{6}$ gates into D3$_{5}$, and the combination of RZZ$_{8}$ and CP$_{9}$ gates leads to the synthesis of D3$_{7}$. The conventional approach is straightforward to implement, but it hinders the potential performance improvements. Note that the conventional approach stops merging the gates when encountering a barrier, RX$_{7}$, during the visiting of the gates for fusion. Conversely, the diagonal detector further checks the property of commuting gates beyond the \emph{barrier} to maximize the size of the gate block. In practice, excessive tensor products of diagonal matrices may still lead to increased optimizer runtime, especially when more qubits are involved. Therefore, a practical upper bound is introduced as a heuristic to mitigate this limitation.

\paragraph{\textbf{Algorithm Design.}} 
As shown in Algorithm~\ref{algo:diagonalDetector}, the diagonal detector restructures quantum circuits using a linked-list-based approach.
Different types of gates are categorized into separate lists: \textit{resList} stores the final optimized circuit, \textit{diagList} collects diagonal gates for subsequent fusion, and \textit{uList} tracks unprocessed gates.
Additionally, a \textit{stopTable} serves as an array and ensures compliance with state vector simulation rules.

Initially, the algorithm processes gates in \textit{gateBlock} sequentially. Gates that are not diagonal are immediately added to the \textit{resList} (Lines \ref{line:not_skip_begin}–\ref{line:not_skip_end}) and bypass further processing.
Diagonal gates, in contrast, are subjected to a set of conditional checks to identify and link mergeable gates for potential fusion.
Accordingly, the detection focus shifts to the subsequent gates (Lines~\ref{line:fIt_begin}-\ref{line:fIt_end}).

If the subsequent gate is not a diagonal gate, instead of terminating the merge immediately, it is preferable to use the set intersection for further evaluation.
In Lines~\ref{line:not_diag_begin}-\ref{line:not_diag_end}, if the required qubits of the gate overlap with those of any previously encountered gates, this information is recorded in the \textit{stopTable}, and the gates are then linked to the \textit{uList}.
This consideration can allow the algorithm to continue detecting subsequent gates.
Specifically, as shown in \figureautorefname{~\ref{fig:diagOpt}}(b), the RZZ$_{2}$ gate can disregard H$_{3}$ and RY$_{4}$, continuing with further detection. The RX$_{7}$ also adheres to the same bypassing rule.
    
In the case of encountering a diagonal gate, the evaluation against the \textit{stopTable} becomes a crucial step.
Should the gate violate the established simulation rules, the process is terminated (Lines~\ref{line:stop_test_begin}-\ref{line:stop_test_end}). 
Subsequently, the gates in the \textit{diagList} that can be merged are processed (Lines~\ref{line:do_diag_begin}-\ref{line:do_diag_end}).
A case in point is the detection of RZZ$_{11}$ following RZZ$_{2}$, which illustrates this condition.
When no stopping conditions are met, diagonal gates continue to be recorded in the \textit{diagList} (Lines~\ref{line:connect_begin}-\ref{line:connect_end}).
This functionality enables algorithm can merge RZZ$_{2}$, RZZ$_{5}$, CP$_{6}$, RZZ$_{8}$, and CP$_{9}$ into D4$_{4}$ successfully.
    
In terms of complexity analysis, despite the presence of two while loops in the diagonal detector, the efficient organization of multiple lists ensures that each gate is accessed only once. As a result, both the time and space complexity are $\mathcal{O}$(\textit{G}), where \textit{G} denotes the total number of gates in the circuit.
\section{Evaluation}
\label{sec:evaluation}
The experimental setup for evaluating the proposed algorithms and our quantum circuit simulator is described in Section~\ref{sec:experimental_env}.
Two types of benchmark sets are considered in our simulation environment: gate-level benchmarks and circuit-level benchmarks.
The experimental results for the entire HPC system are presented in Section~\ref{sec:gateBenchmark} and Section~\ref{sec:appBenchmark}, respectively.
Our performance results are compared with a range of state-of-the-art works, specifically those utilizing circuit optimizers that maintain moderate transpilation costs.
To investigate single-GPU performance, Section~\ref{sec:boosting} provides an analysis of boosting techniques for common circuits, while Section~\ref{sec:ablation} reports ablation studies across various simulators.

\begin{figure*}[htb!]
\captionsetup[subfloat]{farskip=2pt,captionskip=1pt}
\centering
\subfloat[QFT.]{\includegraphics[width=0.475\textwidth]{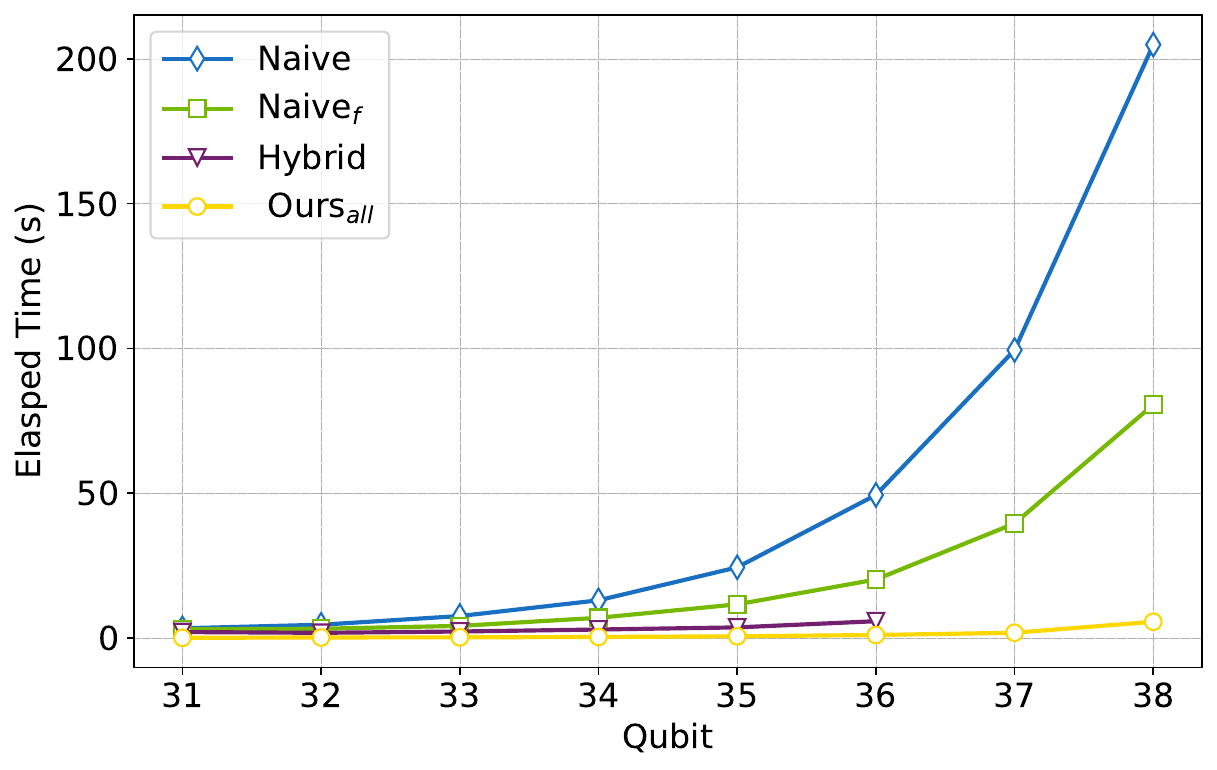}
\label{fig:qft}
}\hfill
\subfloat[5-level fully-connected QAOA.]{\includegraphics[width=0.49\textwidth]{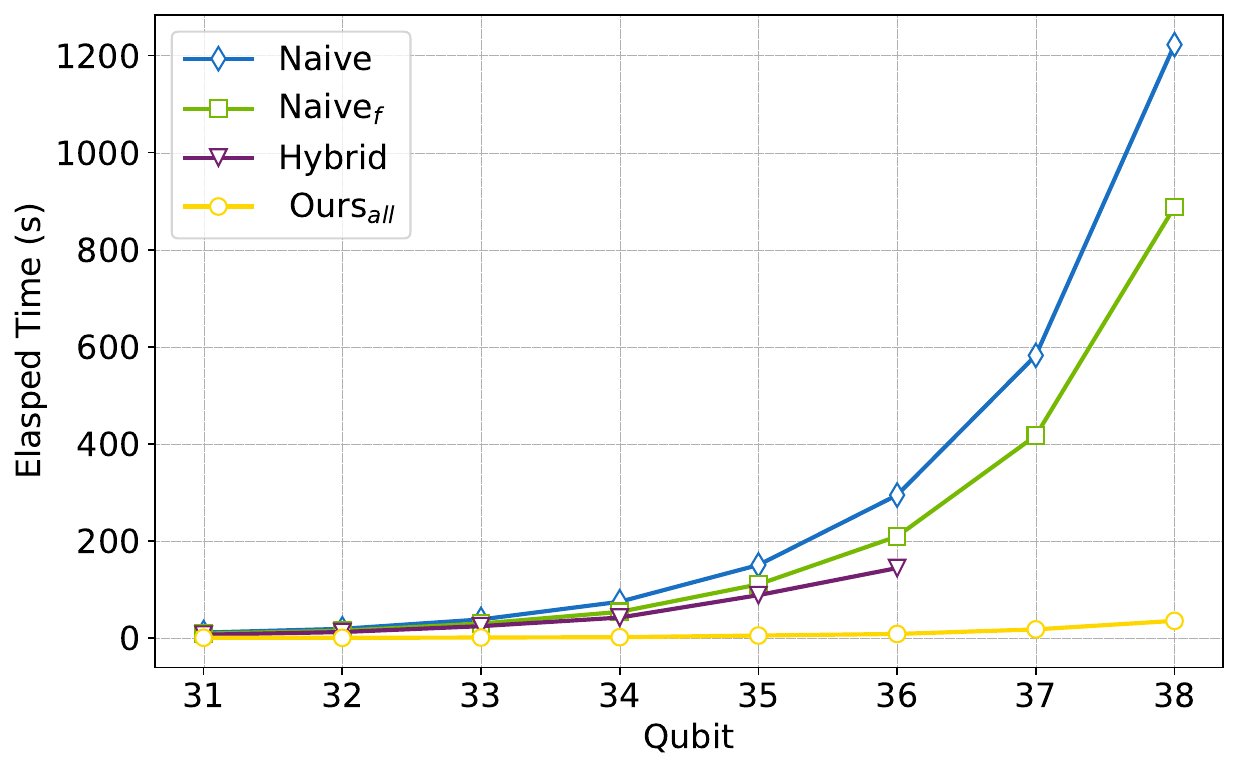}
\label{fig:qaoa}
}
\caption{Elapsed time of the circuit-level benchmarks ranges from 31 to 38 qubits.}
\label{fig:circuit_benchmark}
\end{figure*}

\subsection{Experimental Setup}
\label{sec:experimental_env}
The quantum circuit simulation is performed on an HPC cluster comprising eight NVIDIA H100 workstations~\cite{DGXH100}, each equipped with eight NVIDIA H100 GPUs, resulting in a total GPU memory capacity exceeding 5 TB. Within each workstation, the GPUs are interconnected via NVSwitch, providing an intra-server bandwidth of 900 GB/s. The workstations are connected externally through InfiniBand, with an inter-server bandwidth of 200 GB/s.
The system is powered by two Intel Xeon Platinum 8480 processors and runs on Red Hat Linux 8.5.0 with kernel version 4.18.0.
Given this configuration, simulations of up to 38 qubits can be performed utilizing double-precision floating-point operations.

In our experiments, simulation performance is measured in seconds, utilizing \emph{double-precision} floating-point numbers for data representation across all 64 GPUs.
To effectively demonstrate the impact of the various optimizations, the gate and circuit benchmarks are chosen to represent the complex scenarios.
The gate-level benchmark employs RZZ gates with full connectivity. The circuit-level benchmark focuses on programs with and without the potential for diagonal fusion optimizations, specifically the 5-level fully connected Quantum Approximate Optimization Algorithm (QAOA)~\cite{QAOA} and the Quantum Fourier Transform (QFT)~\cite{QFT}.

Due to the stable performance, the IBM Aer simulator with the cuQuantum~\cite{cuQuantum} backend is used for single-node simulations, while the cusvaer simulator is employed for multi-node simulations.
These simulators without gate fusion serve as the baseline and are referred to as~\emph{Naive}.
HyQuas~\cite{HyQuas}, which achieves the most efficiency under the hybrid scheme in GPU, is treated as the representative method and denoted as~\emph{Hybrid}\footnote{The RZZ gates in the benchmarks are decomposed into equivalent CNOT and RZ gates to account for their absence in HyQuas. Although this decomposition increases the total number of gates, it allows leveraging the advantages of Tensor Cores optimization, thereby enabling performance comparison under this configuration.}.

\subsection{Gate-Level Benchmark}
\label{sec:gateBenchmark}
% To explicitly indicate which optimizations are enabled, as illustrated in Table~\ref{tab:rzz}, the subscript \textit{f} denotes the activation of gate fusion, \textit{b} enables the merge booster, and \textit{all} refers to the activation of all optimizations, including caching blocking, gate fusion, and merge boost.
To explicitly indicate which optimizations are enabled, as illustrated in Table~\ref{tab:rzz_mul}, the subscript \textit{f} denotes the activation of gate fusion, \textit{b} enables the merge booster, and \textit{all} refers to the activation of all optimizations, including caching blocking, gate fusion, and merge booster.
The performance speedup is indicated in parentheses following the simulation time.

Overall, the \textit{Naive} results exhibit exponential growth in runtime as the number of qubits increases.
Applying gate fusion on the \textit{Naive} simulator can yield up to 2× improvement.
\textit{Hybrid} provides notable speedups of up to 6.5× for larger qubit counts but may encounter out-of-memory (OOM) issues on the GPU for 37 qubits and above.
In contrast, our work demonstrates substantial performance gains, with Ours$_{all}$ completing the 38-qubit simulation in under one second and achieving a speedup of over 164× relative to the baseline.
These results indicate that the proposed optimizations can significantly improve computational efficiency while maintaining scalability for large-scale quantum circuit simulations.

\begin{table}[htpb!]
    \caption{Elapsed time of RZZ gates ranging from 30 to 38 qubits (unit: seconds), with speedup shown in parentheses.}
    \label{tab:rzz_mul}
    \centering
    \setlength{\tabcolsep}{3pt} % 增加欄距
    \renewcommand{\arraystretch}{1.05} % 調整行高
    % \begin{tabular}{ccccccc}
    \begin{tabular}{cccccc}
    \hline \specialrule{0em}{1.5pt}{1.5pt}
    % Qubit &  Memory & Memory$_{f}$ & Hybrid & Ours$_{f}$ & Ours$_{b}$ & Ours$_{all}$ \\
    Qubit & Naive & Naive$_{f}$ & Hybrid & Ours$_{b}$ & Ours$_{all}$ \\
    \hline
% 28 &  \phantom{0}0.92 & 0.80 (1.2x) &  0.41 (2.3x) & 0.06 (16.9x) & 0.05 (17.4x) & 0.03 (29.5x) \\
% 29 &  \phantom{0}1.14 & 0.84 (1.4x) &  0.14 (8.1x) & 0.09 (13.1x) & 0.06 (18.5x) & 0.05 (23.6x) \\
% 30 &  \phantom{0}1.84 & 1.13 (1.6x) &  0.28 (6.6x) & 0.13 (13.6x) & 0.10 (19.2x) & 0.05 (35.7x) \\
% 31 &  \phantom{0}2.61 & 1.16 (2.3x) &  0.53 (4.9x) & 0.25 (10.5x) & 0.17 (15.1x) & 0.08 (32.6x) \\
% 32 &  \phantom{0}4.57 & 1.64 (2.8x) &  1.21 (3.8x) & 0.46 (\phantom{0}9.9x) & 0.31 (14.9x) & 0.12 (39.3x) \\
% 33 &  \phantom{0}8.89 & 2.76 (3.2x) &  2.89 (3.1x) & 0.88 (10.1x) & 0.61 (14.5x) & 0.21 (41.9x) \\
% 34 & 17.90 & 5.19 (3.4x) & OOM & 1.79 (10.0x)&  1.27 (14.1x) & 0.41 (43.4x) \\
% 35 & 37.01 & 9.42 (3.9x) & OOM & 3.75 (\phantom{0}9.9x)&  2.60 (14.2x) & 0.82 (45.3x) \\
30&	  \phantom{00}2.68 &	 \phantom{0}2.29 (1.2x) & 	1.55 (1.7x)	& 0.02 (115x)	& 0.02 (157x) \\
31&	  \phantom{00}3.59 &	 \phantom{0}3.16 (1.1x) & 	2.04 (1.8x)	& 0.03 (110x)	& 0.02 (191x) \\
32&	  \phantom{00}3.98 &	 \phantom{0}3.25 (1.2x) & 	1.77 (2.2x)	& 0.06 (67x)	& 0.03 (130x) \\
33&	  \phantom{00}6.08 &	 \phantom{0}4.26 (1.4x) & 	2.23 (2.7x)	& 0.10 (61x)	& 0.04 (157x) \\
34&	 \phantom{0}10.06 &	 \phantom{0}6.26 (1.6x) & 	2.92 (3.4x)	& 0.20 (50x)	& 0.07 (140x) \\
35&	 \phantom{0}18.69 &	10.37 (1.8x) & 	3.65 (5.1x)	& 0.39 (48x)	& 0.11 (164x) \\
36&	 \phantom{0}37.83 &	18.53 (2.0x) & 	5.79 (6.5x)	& 0.80 (47x)	& 0.22 (169x) \\
37&	 \phantom{0}75.45 &	37.37 (2.0x) & 	    OOM 	& 1.69 (45x)	& 0.45 (166x) \\
38&	151.52 &	74.24 (2.0x) &  	    OOM 	& 3.50 (43x)	& 0.93 (164x) \\
    \hline
    \end{tabular}
\vspace{-0.3cm}
\end{table}

\subsection{Circuit-Level Benchmark}
\label{sec:appBenchmark}

% \figureautorefname{~\ref{fig:circuit_benchmark}(a)} and \figureautorefname{~\ref{fig:circuit_benchmark}(b)} present the simulation times for QFT and QAOA, respectively, with qubits ranging from 27 to 35.
% Due to recurring out-of-memory issues of the hybrid method, the 34-qubit and 35-qubit simulations are not specifically addressed and discussed here for the sake of completeness.

\figureautorefname{~\ref{fig:circuit_benchmark}} presents the simulation times for QFT and QAOA, respectively, with qubits ranging from 31 to 38.
Due to recurring out-of-memory issues of the hybrid method, the 37-qubit and 38-qubit simulations are not specifically addressed and discussed here.

In the case of the QFT circuit, \textit{Naive}$_{f}$ achieves a 2.4× performance improvement over \textit{Naive} when gate fusion is enabled. In the 36-qubit simulation, \textit{Hybrid} attains speedups of 8.5× and 3.5× relative to \textit{Naive} and \textit{Naive}$_{f}$, respectively.
Our framework can outperform the others, achieving a remarkable 34× speedup over \textit{Naive} and operating 5.8× faster than \textit{Hybrid}.
For the QAOA simulation, \textit{Hybrid} can consistently outperform \textit{Naive} and \textit{Naive}$_{f}$ across all scenarios.
With a robust implementation ensuring efficient cross-node communication, our simulator delivers superior performance across all tested scenarios, achieving speedups of up to 34 times.

% \ccw{In the case of the QFT circuit, \textit{Naive}$_{f}$ achieves a 1.6× performance improvement over \textit{Naive} when gate fusion is enabled. In the 33-qubit simulation, \textit{Hybrid} attains speedups of 4.3× and 2× relative to \textit{Naive} and \textit{Naive}$_{f}$, respectively.
% Despite the limited total number of diagonal gates in the QFT circuit, our framework outperforms the others, achieving a 23× speedup over \textit{Naive} and operating 5.5× faster than \textit{Hybrid}.}

% For the QAOA simulation, \textit{Hybrid} can outperform \textit{Naive} and \textit{Naive}$_{f}$ when processing up to 32 qubits; however, an unexpected performance drop occurs at 33 qubits.
% In comparison with the subsequent single-device ablation experiments, \ccw{the impact of introducing third-party libraries for cross-device simulations may require further performance analysis.}
% In contrast, our simulator consistently maintains a leading performance across all tested cases, achieving speedups of up to 12.7x.

\subsection{Analysis of Boosting Optimization} \label{sec:boosting}
Upon completing large-scale, multinode simulations, it is essential to understand the performance impact of various optimizations on individual hardware units.
For a 31-qubit simulation on a single device, Table~\ref{tab:version_comparison} reports the per-gate execution time of different simulation approaches across benchmark circuits drawn from various application domains~\cite{qibojit_benchmarks}.
The baseline \textit{Naive} exhibits a consistent per-gate cost of approximately 33 to 43 milliseconds.
% This is primarily because simulating an \textit{N}-qubit circuit requires each gate to perform $2^{N}$ memory accesses to update the entire state.
%quantum state.

Depending on the quantum circuit, enabling gate fusion typically yields a 3- to 5-fold performance improvement, but it also increases the variability of the average per-gate execution time.
The cache blocking mechanism consistently reduces the per-gate cost to below 5 microseconds, resulting in a performance improvement of up to 10 times across all benchmarks. 
This indicates that modifying the simulation scheme should be prioritized over general gate fusion.
With boosting optimization enabled (Ours$_{b}$), further improvements are observed, yielding speedups ranging from 1.4× to 3.5×.
The Hidden Shift (HS) and Supremacy Circuit (SC) circuits achieve up to 2.9× and 3.5× speedups, respectively, as the entire circuit only requires full entanglement at later stages.
These results indicate that integrating the booster also provides substantial performance benefits compared to gate fusion and cache blocking.

\begin{table}[htb!]
    \centering
    \caption{Comparison of per-gate performance and speedup on 31-qubit circuits (unit: milliseconds).}
    \label{tab:version_comparison}
    \setlength{\tabcolsep}{2mm} % 調整欄距
    \renewcommand{\arraystretch}{1} % 調整行高
    \begin{tabular}{|c|c|c|c|c|c|}
        \hline
        Circuit & Naive & Naive$_{f}$ & Ours & Ours$_{b}$ & Speedup$_{b}$ \\
        \hline
        BV~\cite{bv}     & 43.21 & 12.27 & 4.76 & 3.51 & 1.4x \\ 
        HS~\cite{hs}     & 42.95 & \phantom{0}7.19  & 3.78 & 1.31 & 2.9x \\
        QAOA~\cite{QAOA} & 41.99 & 12.60 & 4.26 & 2.35 & 1.8x \\
        QFT~\cite{QFT}   & 33.94 & 11.74 & 3.44 & 1.26 & 2.7x \\
        QV~\cite{qv}     & 41.25 & \phantom{0}4.33 & 3.39 & 2.21 & 1.5x \\
        SC~\cite{sc}     & 41.30 & \phantom{0}6.93 & 3.18 & 0.91 & 3.5x \\
        VC~\cite{vc}     & 41.39 & \phantom{0}8.31 & 3.47 & 1.28 & 2.7x \\
        \hline
    \end{tabular}
    \vspace{-0.3cm}
\end{table}

\subsection{Ablation Experiment}
\label{sec:ablation}
To systematically assess the impact of various optimizations on simulation performance in a fundamental setup, we conducted ablation experiments.
The evaluation was conducted on the most complex 30-qubit QAOA circuit using a single NVIDIA H100 GPU, with a particular emphasis on our optimization techniques.

In the simulation scheme without the caching block technique, there is generally minimal variation in execution times between the different simulators.
QuEST is slightly faster due to its specialized implementation for the typical gates.
% When gate fusion is utilized, a speedup of at least 3x is achieved.
When gate fusion is utilized, the Aer Simulator, using cuTrust and cuQuantum as the backends, achieves a speedup of at least threefold in single-machine.
A specialized simulator~\cite{Lin_2024} achieves a speedup of more than 9.7 times by utilizing preprocessed and optimized code for diagonal gates.
HyQuas employs its hybrid approach and leverages the cuBLAS library, resulting in a 5.3x optimization.
It is worth noting that for programs exhibiting great potential for diagonal fusion optimization, such as QAOA, a simulator specifically optimized for diagonal fusion optimization is faster than a highly optimized one. 

% 單純為了在 overleaf 好看。
% In our evaluation under standard operating conditions, our simulator consistently outperforms the performance of all existing cost-based fusion techniques.
% When the merge booster optimization is activated, Ours$_{b}$ delivers an additional speedup.
% However, when only diagonal fusion is activated, Ours$_{f}$ has a slightly lower performance than that of the specialized simulator.
% \format{This subtle performance gap can be attributed to the restricted fusion size to fit the processor cache, which is necessary to maintain the efficiency of the tensor product during the optimization stage for the circuit.
% Lastly, when we enable all optimization strategies, our approach demonstrates remarkable compatibility and a notable performance improvement of 29.6 times.}

Under typical conditions, our simulator outperforms all cost-based fusion approaches.
When the merge booster optimization is enabled, Ours$_{b}$ delivers an additional speedup.
However, when only diagonal fusion is activated, Ours$_{f}$ has a slightly lower performance than that of the specialized simulator.
This discrepancy can be attributed to the restricted fusion size to fit the processor cache, which is necessary to maintain the efficiency of the tensor product during the optimization step for the circuit.
Lastly, when all optimizations are enabled, our approach remains fully compatible and demonstrates a notable performance improvement of 29.6 times.
\begin{table}[htb!]
    \caption{Elapsed time of 30-qubit fully-connected QAOA with different optimizations (unit: seconds).}
    \label{tab:ablation}
    \centering
    \setlength{\tabcolsep}{1.5mm}
    \renewcommand{\arraystretch}{1} % 調整行高
    \begin{tabular}{|c|c|c|c|c|c|}
    \hline %\specialrule{0em}{1.5pt}{1.5pt}
     Simulator & Cache & Fusion & Boosting & Time & Speedup \\
    \hline
        cuQuantum & No & No & No & 29.41 & - \\ 
        % \ccw{cuQuantum$_{d}$} & No & No & No & 64.02 & - \\ 
        AerSimulator & No & No & No & 28.04 & \phantom{0}1.0x \\ 
        QuEST & No & No & No & 26.68 & \phantom{0}1.1x \\ 
    \hline
        % \ccw{cuQuantum$_{df}$} & No & Yes & No & 1.97 & - \\ 
        cuQuantum$_{f}$ & No & Yes & No & \phantom{0}9.09 & \phantom{0}3.2x \\ 
        AerSimulator$_{f}$ & No & Yes & No & \phantom{0}8.86 & \phantom{0}3.3x \\ 
        Lin et al$_{f}$~\cite{Lin_2024} & No & Yes & No & \phantom{0}3.03 & \phantom{0}9.7x \\ 
    \hline
        \raisebox{-0.5ex}{HyQuas} & \multicolumn{2}{c|}{\raisebox{-0.5ex}{Hybrid}} & \raisebox{-0.5ex}{No} & \raisebox{-0.5ex}{\phantom{0}5.57} & \raisebox{-0.5ex}{\phantom{0}5.3x} \\ [0.3em]
    \hline
        Ours & Yes & No & No & \phantom{0}5.02 & \phantom{0}5.8x \\ 
        Ours$_{b}$ & Yes & No & Yes & \phantom{0}2.76 & 10.6x \\ 
        Ours$_{f}$ & Yes & Yes & No & \phantom{0}3.07 & \phantom{0}9.6x \\ 
        Ours$_{all}$ & Yes & Yes & Yes & \phantom{0}0.99 & 29.6x \\ 
    \hline
    \end{tabular}
    \vspace{-0.2cm}
\end{table}

% \begin{table}[tb!]
%     \caption{The elapsed time of 30-qubit QAOA with different optimizations (unit: seconds). The speedup is labeled as SU.}
%     \label{tab:gpu}
%     \setlength{\tabcolsep}{1.8mm}{
%     \begin{tabular}{|c|c|c|c|c|c|}
%     \hline %\specialrule{0em}{1.5pt}{1.5pt}
%      Simulator & Caching & Fusion & Boost & Time & SU \\
%     \hline
%         cuQuantum & No & No & No & 29.41 & - \\
%         AerSimulator & No & No & No & 28.04 & \phantom{0}1.0x \\
%         QuEST & No & No & No & 26.68 & \phantom{0}1.1x \\
%     \hline
%         cuQuantum$_{f}$ & No & Yes & No & \phantom{0}9.09 & \phantom{0}3.2x \\
%         AerSimulator$_{f}$ & No & Yes & No & \phantom{0}8.86 & \phantom{0}3.3x \\
%         Lin et al$_{f}$~\cite{Lin_2024} & No & Yes & No & \phantom{0}3.03 & \phantom{0}9.7x \\
%     \hline
%         HyQuas$_{f}$ & Yes & Yes & No & \phantom{0}5.57 & \phantom{0}5.3x \\        
%     \hline
%         Ours & Yes & No & No & \phantom{0}5.02 & \phantom{0}5.8x \\
%         Ours$_{b}$ & Yes & No & Yes & \phantom{0}2.76 & 10.6x \\
%         Ours$_{f}$ & Yes & Yes & No & \phantom{0}3.07 & \phantom{0}9.6x \\
%         Ours$_{all}$ & Yes & Yes & Yes & \phantom{0}0.99 & 29.6x \\
%     \hline
%     \end{tabular}
%     }
% \vspace{-0.3cm}
% \end{table}
\section{Conclusion}
\label{sec:conclusion}
This work aims to advance a full-state quantum circuit simulation framework, inspired by our empirical findings of two key circuit properties—\emph{inter-gate block entanglement-free} behavior and \emph{commuting gates}—as well as a practical limitation in using existing third-party libraries to handle fused gate operations.

To address these findings, the methodologies (\emph{merge booster} and \emph{diagonal detector}) are proposed and integrated into a quantum circuit optimizer, a common preprocessing step for accelerating quantum circuit simulation prior to execution.
The developed simulator outperforms state-of-the-art systems by exploiting data locality and computational efficiency through the proposed \emph{merge booster} and \emph{diagonal detector}.
On a single device, performance can exceed the baseline by more than ten times across all benchmarks.
By extending the MPI implementation, our proposed approaches still achieve a speedup of more than one order of magnitude.

Owing to the strong extensibility of our simulator, future work will explore incorporating the proposed algorithms into existing simulators to further broaden the impact of the methodologies.

% The performance analysis and ablation experiments further validate the feasibility of the proposed approach.
% Designed with polynomial time and space complexity, these methodologies incur only limited processing overhead while delivering substantial speedups.
% Owing to the strong extensibility of our simulator, future work will explore incorporating the proposed algorithms into existing simulators to further broaden the impact of the methodologies.

\bibliographystyle{ACM-Reference-Format}
\bibliography{sample-base}

\end{document}